\documentstyle[12pt,emulateapj]{article}

\lefthead{Nitta}
\righthead{Reconnection rate in the self-similar evolution model}

\begin{document}

\title{Outflow structure and reconnection rate of the self-similar evolution model of fast magnetic reconnection}

\author{Shin-ya Nitta \altaffilmark{1} \altaffilmark{2}}
\affil{The Coordination Center for Research and Education, 
The Graduate University for Advanced Studies (SOKENDAI), 
Shonan Village, Hayama, Kanagawa, 240-0193 Japan}
\authoremail{snitta@koryuw02.soken.ac.jp}

\altaffiltext{1}{Division of Theoretical Astrophysics, National Astronomical Observatory of Japan, Osawa 2-21-1, Mitaka 181-8588, Japan}

\altaffiltext{2}{Department of Astronomical Science, The Graduate University for Advanced Studies, Osawa 2-21-1, Mitaka 181-8588, Japan}

\begin{abstract}

In order to understand the nature of magnetic reconnection in ``free space'' which is free from any influence of external circumstances, I have studied the structure of spontaneous reconnection outflow using a shock tube approximation. The reconnection system of this case continues to expand self-similarly. This work aims 1) to solve the structure of reconnection outflow and 2) to clarify the determination mechanism of reconnection rate of the ``self-similar evolution model'' of fast reconnection. 

Many cases of reconnection in astrophysical phenomena are characterized by a huge dynamic range of expansion in size ($\sim 10^7$ for typical solar flares). Although such reconnection is intrinsically time dependent, a specialized model underlying the situation has not been established yet. 

The theoretical contribution of this paper is in obtaining a solution for outflow structure which is absent in our previous papers proposing the above new model. The outflow has a shock tube-like structure, i.e., forward slow shock, reverse fast shock and contact discontinuity between them. By solving the structure in a sufficiently wide range of plasma-$\beta$: $0.001 \leq \beta \leq 100$, we obtain an almost constant reconnection rate ($\sim 0.05$: this value is the maximum for spontaneous reconnection and is consistent with previous models) and a boundary value along the edge of the outflow (good agreement with our simulation result) which is important to solve the inflow region. Note that everything, including the reconnection rate, is spontaneously determined by the reconnection system itself in our model. 

\end{abstract}

\section{Introduction}
\label{sec:Int}

It is widely accepted that magnetic reconnection very commonly plays an important role as a powerful energy converter in astrophysical plasma systems. However, there still remain many open questions, not only regarding the microscopic physics of the anomalous resistivity, but also the macroscopic magnetohydrodynamical (MHD) structure. In this work, our attention is focused on the macroscopic evolutionary properties of magnetic reconnection, especially on the structure of the reconnection outflow.

We consider reconnection in an anti-parallel magnetic configuration called a ``current sheet system''. In this system, two similarly uniform magnetized regions are set in contact, divided by a boundary. We assume that the directions of the magnetic field on both sides are anti-parallel to clarify the essence of the problem. In this case, the boundary carries a strong electric current, hence, we call it the current sheet. If resistivity is enhanced in this current sheet, it plays an important role in energy conversion from the magnetic form to others. In resistive plasmas, there are two fundamental processes of magnetic energy conversion. One is magnetic diffusion, and the other is magnetic reconnection. 

The best known process of energy conversion in resistive plasmas from magnetic to other forms is Ohmic dissipation (magnetic diffusion). We must note, however, that plasmas are highly conductive in most astrophysical problems. Since the resistivity is very small, energy conversion by global magnetic diffusion takes a very long time, and is not applicable to many astrophysical phenomena with very violent energy releases, e.g., solar flares. Even in such a case, the majority believe that magnetic reconnection can convert the magnetic energy very quickly. This is why we must study magnetic reconnection. 

Such energy conversion in the current sheet system is very important in many astrophysical plasma systems. The most famous example is the relation between solar activity and geomagnetospheric activity, called solar flares and geomagnetospheric substorms. In this case, at least three current sheet systems are included. The first one is in the solar corona. The second and third are on the day-side and the night-side of the geomagnetosphere, respectively. Recent observations and numerical simulations suggest that similar phenomena, seemingly activated by magnetic reconnection, are universal, e.g., flares in accretion disks of young stellar objects (YSOs, see Koyama et al. 1994, Hayashi et al. 1999), galactic ridge X-ray emissions (GRXE, see Koyama et al. 1986, Tanuma et al. 1999). 

We must note that many cases of actual magnetic reconnection in astrophysical systems usually grow over a huge dynamic range in their spatial dimensions. For example, the initial scale of the reconnection system can be defined by the initial current sheet thickness, but this is too small to be observed in typical solar flares. We do not have any convincing estimate of the scale, but if we estimate it to be of the order of the ion Larmor radius, it is extremely small ($\sim 10^0$ [m] in the solar corona). Finally, the reconnection system develops to a scale of the order of the initial curvature radius of the magnetic field lines ($\sim 10^7$ [m] $\sim$ 1.5\% of the solar radius for typical solar flares). The dynamic range of the spatial scale is obviously huge ($\sim 10^7$ for solar flares). For geomagnetospheric substorms, their dynamic range of growth is also large ($\sim 10^4$ for substorms). Such a very wide dynamic range of growth suggests that the evolution of the magnetic reconnection should be treated as a development in ``free space'', and that external circumstances do not affect the evolutionary process of magnetic reconnection, at least at the expanding stage just after the onset of reconnection. 

Even if a system is not completely free from the influence of external circumstances, we can approximately treat it as a spontaneously evolving system if the evolution timescale (which is estimated as Alfv\'{e}n transit time $\tau_A$ for the final system scale because the spontaneous expansion speed of the reconnection system is equivalent to the fast-mode propagation speed, see Nitta et al. 2001; hereafter paper 1) is much smaller than the timescale imposed by the external circumstances (e.g., convection timescale). For typical substorms, the evolution timescale of the reconnection is of the order of seconds or minutes while the convection timescale to compress the current sheet is of the order of hours. In such cases, the external circumstances simply play the role of triggering the onset of reconnection and the evolution itself is approximately free from the influence of the external circumstances. Though, of course, exceptional cases in which the external circumstances intrinsically influence the evolution can arise in particular situations (e.g., cases in which very fast convection drives the reconnection), the author believes that it is worth establishing a new model applicable to the cases which are free from external influences.

However, no reconnection model evolving in a free space has been studied. We should note that most previous theoretical and numerical works on reconnection treated it as a boundary problem strongly influenced by external circumstances. 

In actual numerical studies, there is a serious and inevitable difficulty: For magnetic reconnection to be properly studied, the thickness of the current sheet must be sufficiently resolved by simulation mesh size. Usually, the thickness of the current sheet is much smaller than the entire system. Hence, in previous works, we have been forced to cut a finite volume out of actual large-scale reconnection systems for numerical studies. 

Of course, we wish to realize that the boundary conditions of this type of finite simulation box reproduce the evolution in unbounded space. However, we should note that, in actual simulations, the boundary of such a simulation box necessarily affects the evolution inside the box, even if we use so-called ``free'' boundary conditions. This has an adverse influence on the evolutionary process of the numerically simulated magnetic reconnection: When the reconnection proceeds and physical signals propagating from the inner region cross the boundaries of the simulation box, the subsequent evolution is necessarily affected by the boundary conditions. This is because the information propagated through the boundary is completely lost and such an artificially cut-out simulation box will never receive the proper response of the outer regions. Additionally, numerical and unphysical signals emitted from the boundary may disturb the evolution. Such artificially affected evolution is obviously unnatural, and the resultant stationary state should differ from actual reconnection occurring in a free space. 

Even in previous numerical studies aimed at clarification of the time evolution of reconnection (for example, series of studies originated by Ugai \& Tsuda 1977 or Sato \& Hayashi 1979), the evolution could not be followed for a long time. This is mainly owing to restriction of the size of the simulation box, hence, application of the results was limited to spatial scales typically, say, a hundred times the spatial scale of the diffusion region. We are interested here in an evolutionary process in free space without any influence of external circumstances. In such a system, the evolution and resultant structure would be quite different from these previous numerical models. 

The same problem also appears in previous theoretical works. We must realize that, in many cases, astrophysical magnetic reconnection is essentially a non-stationary process because it grows in a huge spatial dynamic range. Most of these works, however, for mathematical simplicity, treat a stationary state of the reconnection in a finite volume (for example, see Priest \& Forbes 1986). These are obviously boundary problems and solutions must be influenced by boundary conditions. The problem is that we do not know how we should set the boundary condition in order to simulate external influences. In general, it is impossible to set. Hence, the situations argued in previous works are rather artificial and unnatural in many cases of astrophysical application. 

As will be discussed later (see section \ref{sec:stat}), if we cut the central region out from the entire expanding system, the central region tends to be a stationary state and is very similar to the well-known Petschek model, i.e., the system is characterized by the fast-mode rarefaction dominated inflow and the figure-X-shaped slow shock. The author thinks that stationary solutions are meaningful as the interior solution of the entire evolving system in free-space though such stationary models are frequently treated as externally ``driven'' processes by boundary conditions. 

We demonstrated the above process of self-similar evolution model of fast reconnection by numerical simulation (see paper 1) and semi-analytic study (see Nitta et al. 2002; hereafter paper 2). Let us overview such an evolutionary process in free-space. 

We suppose a two-dimensional equilibrium state with an anti-parallel magnetic field distribution, as in the Harris solution. When magnetic diffusion takes place in the current sheet by some localized resistivity, magnetic reconnection will occur, and a pair of reconnection jets is ejected along the current sheet. This causes a decrease in total pressure near the reconnection point. Such information propagates outward as a rarefaction wave. In a low-$\beta$ plasma ($\beta \ll 1$ in the region very distant from the current sheet [asymptotic region]; as typically encountered in astrophysical problems), the propagation speed of the fast-magnetosonic wave is isotropic, and is much larger than that of other wave modes. Thus, information about the decreasing total pressure propagates almost isotropically as a fast-mode rarefaction wave (hereafter FRW) with a speed almost equal to the Alfv\'{e}n speed $V_{A0}$ in the asymptotic region. Hence, the wave front of the FRW (hereafter FRWF) has a cylindrical shape except near the point where the FRWF intersects with the current sheet. When the FRWF sufficiently expands, the initial thickness of the current sheet becomes negligible compared with the system size $V_{A0} t$, where $t$ is the time from the onset of reconnection. In such a case, there is only one characteristic scale, i.e., the radius of the FRWF ($V_{A0} t$), which linearly increases as time proceeds. This is just the condition for self-similar growth.

In our semi-analytic work (paper 2), the boundary condition along the edge of the outflow is artificially imposed approximating the result of our numerical simulation in paper 1. We set the boundary condition at $y=0$ for simplicity because the outflow is very thin. This boundary condition denotes the junction condition to the reconnection outflow. Hence, an important question ``how is this boundary condition spontaneously determined by the reconnection system'' still remains. Our motivation of this work is to clarify the mechanism to determine this boundary condition.

The boundary condition at $y=0$ determines the inflow region. Hence, the reconnection rate which is usually denoted by the normalized inflow speed toward the diffusion region is also determined. We must note that, in our self-similar evolution model, everything including the reconnection rate is spontaneously determined by the reconnection system itself. We try to clarify how the system regulates the reconnection.  

This paper is organized as follows. We state a model of reconnection outflow as a kind of shock tube problem in section \ref{sec:model}. Basic equations for outflow structure and the numerical procedure to solve the equations are listed in section \ref {sec:b-eqs}. Properties of the result, e.g., the reconnection rate and boundary condition along the slow shock are shown in section \ref{eq:res}. In section \ref{sec:sum-dis}, we summarize our study and discuss spontaneous structure formation in the reconnection outflow.

\section{Model of reconnection outflow}
\label{sec:model}

We present a schematic picture of the reconnection outflow in our self-similar evolution model (see figure \ref{fig:sch}). A similar structure of reconnection outflow is shown in figure 9 of Ugai 1999 as a simulation result. Because of symmetry with respect to the $x-$ and $y-$axes, we treat only the region $x,\ y>0$. The outflow is composed of two different plasmas which have different origins. The rear-half region ($x < x_c$: reconnection jet) is filled with reconnected plasma coming from outside the current sheet (inflow region). The front-half region ($x > x_c$: plasmoid) is filled with the original current sheet plasma. These two regions are divided by a contact discontinuity located at $x=x_c$.

The entire outflow is surrounded by a slow shock which has a complicated `crab-hand' shape. The Petschek-like figure-X shaped slow shock (oblique shock) is elongated from the diffusion region with a slight opening angle $\theta$. There is a reverse fast shock (almost perpendicular shock) inside the reconnection jet ($x=x_f$). In front of the plasmoid, a forward V-shaped slow shock (oblique shock) forms. The opening angle and the locus (crossing point with $x-$axis) are $\phi$ and $x_s$, respectively. 

The entire structure, including several discontinuities, is analogous to the one dimensional shock tube problem. The forward shock (V-shaped slow shock) and the reverse shock (fast shock) are formed by the collision of the reconnection jet and the original current sheet plasma, and propagate in both directions. Between these two shocks, the contact discontinuity forms. We approximate this reconnection outflow as a quasi 1-D problem in order to solve it analytically. Such an approximation may be valid near the $x-$ and $y-$ axes, because the system is symmetric with respect to these axes. 

We focus our attention on the quasi one dimensional problem of the figure-L shaped region near the $x-$ and $y-$axes (We treat the region $x, y =finite \gg D$ where $D$ is the initial current sheet thickness. Note that this region apparently tends to coincide with the $x-$ and $y-$axes in the self-similar stage which is a very late stage from the onset when we observe the evolution in a zoom-out coordinate [see paper 1].). Each region between two neighboring discontinuities is approximated to be uniform. We note about the up-stream region p just above the X-shaped slow shock that the region between the slow shock and the separatrix field line (X-shaped field line reaching the X-point) is also approximated to be uniform and the $x-$component of the velocity vanishes. In this region, each reconnected field line has an almost straight shape and crosses the X-slow shock while each field line has a hyperbolic shape in the region above the separatrix field line. 

Region 3, between the contact discontinuity and the V-shaped slow shock, looks to be non uniform. Since we do not know how we can solve the two dimensional structure of this region analytically, we approximate this region to be uniform. The effect of this rough treatment is estimated in section \ref{sec:bc}. 

In the vicinity of the diffusion region of self-similar reconnection, the solution is quasi stationary. Hence, the inflow speed toward the diffusion region must equal the magnetic diffusion speed. We here introduce the main assumption of this work: The magnetic diffusion speed is fully adjustable with inflow speed into the diffusion region. This is possible by tuning the thickness of the diffusion region if the resistivity is sufficiently large. Detailed discussion on the validity of this assumption is made in section \ref{sec:Rey-indep}. We note that the reconnection system in free space has only two parameters: the magnetic Reynolds number and the plasma-$\beta$ at the asymptotic region (see paper 1). If the diffusion speed is adjustable, the inflow speed is not restricted by the diffusion speed. Hence, the reconnection may not depend on the magnetic Reynolds number (Yokoyama \& Shibata 1994). Thus, our attention will be focused on the dependence on the plasma-$\beta$. We wish to emphasize that the resistivity does not regulate the system in such a situation (It may play a passive role) though electric resistivity is necessary in order to reconnect the magnetic field lines. We now wonder what mechanism does regulate the system. In order to answer this question, we consider the following shock tube-like model. We can treat the main part of such a problem by ideal (non-resistive) MHD.

The quantities denoting the initial uniform equilibrium at the asymptotic region are gas pressure $P_0$, mass density $\rho_0$ and magnetic field strength $B_0$. Plasma-$\beta$ at the asymptotic region is defined as $\beta \equiv P_0/(B_0^2/[2 \mu])$ where $\mu$ is magnetic permeability of vacuum. In the rest of this paper, we use the normalization of physical quantities as in paper 2. We define units for each dimension as follows: (unit of velocity)$=V_{A0} \equiv B_0/\sqrt{\mu \rho_0}$ (Alfv\'{e}n speed at the asymptotic region), (unit of the length)$=V_{A0} t$ where $t$ is the time from the onset of reconnection, (unit of mass density)$=\rho_0$, (unit of magnetic field)$=B_0$, (unit of pressure)$=\beta/2 \cdot \rho_0 V_{A0}^2$.  

We set the 1-D shock tube-like problem as follows (see figure \ref{fig:sch}). The system has 22 unknown quantities: $P_p$, $\rho_p$, $v_{yp}$, $B_{xp}$, $B_{yp}$, $\theta$, $P_1$, $\rho_1$, $v_1$, $B_1$, $x_f$, $P_2$, $\rho_2$, $v_2$, $B_2$, $P_3$, $\rho_3$, $v_{y3}$, $B_{x3}$, $B_{y3}$, $\phi$ and $x_s$ where $P_*$, $\rho_*$, $v_*$, $B_*$ denote the pressure, density, velocity, and magnetic field, respectively (Note $x_c=v_{x3}=v_2$ because no mass flux passes through the contact discontinuity). The suffices $p$, 1, 2 or 3 denote the region divided by the discontinuities. The suffices $x$ or $y$ denote the vector components. Quantities $\theta$, $x_f$, $\phi$ and $x_s$ denote the inclination of the X-shaped slow shock, the locus of the fast shock, the inclination of the V-shaped slow shock and the locus of the V-shaped slow shock (crossing point with $x-$axis), respectively. 

These unknowns should be related to each other via conditions coming from the integrated form of conservation laws (i.e., the Rankine-Hugoniot [R-H] conditions) or other relations. The set of relations is listed in the next section. 

When $\beta>0.8$, we must change the model because the reverse fast shock does not form in the reconnection outflow. The criterion of $\beta$ whether the reverse fast shock forms or not is in the range $\beta=0.7-0.8$. Though we did not precisely estimate the critical value of $\beta$, we can fairly match the solutions of these two schemes around $\beta \sim 0.75$. For the case including no fast shock, we remove $x_f$ from the unknowns and set $Q_1=Q_2$ where $Q_*$ represents the physical quantities at regions 1 or 2. According to this alteration, the coupled equations of the unknowns are also altered, but we abbreviate their detailed forms because this alteration is trivial.

\section{Basic equations and numerical approach}
\label{sec:b-eqs}

According to the above model of the reconnection outflow, we obtain the following 22 equations for 22 unknown quantities (Detailed forms of each equation are listed in the appendix). \\

\noindent
1)relations between the pre-X-slow shock and the asymptotic region \\
Frozen-in condition [eq1]\\
Polytropic relation [eq2]\\

\noindent
2)R-H jump conditions at the X-slow shock\\
Pressure jump [eq3]\\
Density jump [eq4]\\
Velocity jump (parallel comp.) [eq5]\\
Velocity jump (perpendicular comp.) [eq6]\\
Magnetic field jump (parallel comp.) [eq7]\\
Magnetic field jump (perpendicular comp.) [eq8]\\

\noindent
3)R-H jump conditions at the reverse fast shock\\
Pressure jump [eq9]\\
Density jump [eq10]\\
Velocity jump [eq11]\\
Magnetic field jump [eq12]\\

\noindent
4)Magnetic flux conservation at the X-point [eq13]\\

\noindent
5)Force balance at the contact discontinuity [eq14]\\

\noindent
6)R-H jump conditions at the forward V-slow shock\\
Pressure jump [eq15]\\
Density jump [eq16]\\
Velocity jump (parallel comp.) [eq17]\\
Velocity jump (perpendicular comp.) [eq18]\\
Magnetic field jump (parallel comp.) [eq19]\\
Magnetic field jump (perpendicular comp.) [eq20]\\

\noindent
7)Boundary Condition at the tip of the outflow [eq21]\\

\noindent
8)Magnetic flux conservation all over the outflow [eq22]\\

We can solve the majority of these equations by hand, and by substituting the solutions into other equations, we can reduce equations. Finally, nine equations 3, 4, 5, 6, 7, 11, 14, 21 and 22 remain as complicated nonlinear coupled equations for nine unknowns $v_1$, $B_{xp}$, $B_{yp}$, $\theta$, $P_1$, $\rho_1$, $x_f$, $\phi$ and $x_s$.

We solve these coupled nine equations by an iterative method (Newton-Raphson). In order to find the solution of nonlinear coupled equations, in general, a precise initial guess of the unknowns is required. Unfortunately, we cannot guess the solution so easily, so we adopt the following primitive method to find a good initial guess of the unknowns. 

 We estimate them from the result of our numerical simulation in paper 1 for the case $\beta=0.2$ in the asymptotic region. Of course, the numerical result is contaminated by numerical noise, and so we can treat our simulation result as simply a hint.  

Each equation can be reduced to a form (LHS of eq.)=0. When we substitute the initial guess of physical quantities from our simulation result for $\beta=0.2$ into the unknown quantities, the left hand side does not vanish because these are not the true solution. By defining the residue as the sum of absolute values of the left-hand side of each equation, we can plot a graph of the residue as a function of nine unknown quantities. Then we improve the guess of each unknown to diminish the residue. This process will be iteratively continued by hand until the residue settles to a value as small as possible (until the residue becomes two orders of magnitude smaller than residue for the simulation result). This is a very time-consuming procedure. Thus we can reach a good initial guess for the unknowns in order to use the Newton-Raphson method for $\beta=0.2$. 

First, we start with the case $\beta=0.2$. Once we find a converged solution of the Newton-Raphson procedure for the case $\beta=0.2$, we treat it as the initial guess for the case, e.g., $\beta=0.11$ or $\beta=0.21$, and we have successively obtained the solutions for a range of $0.001 \leq \beta \leq 0.7$. 

For the case $\beta>0.8$, we replace the numerical procedure with another one for the case of no reverse fast shock. When $\beta=0.8$, we adopt the solution for $\beta=0.7$ as the initial guess. Thus, we can continue the converging procedure to the case of very large $\beta$ and obtain the result for $\beta \leq 100$.

\section{Result}
\label{eq:res}
\subsection{Reconnection rate}
We investigated the structure of the reconnection outflow by using a shock tube approximation. The main issues of this work are 1) reconnection rate of the self-similar evolution model and 2) boundary condition along the edge of the outflow. 

Most interest is focused on the reconnection rate. Let us show the $\beta$-dependence of the reconnection rate $R$. The reconnection rate is defined by the inflow magnetic flux per unit time: $R \equiv -V_{yp} B_{xp}$ in our dimensionless form. Figure \ref{fig:rec-rate} clearly shows that the reconnection rate is almost constant in a very wide range of $\beta$. The reconnection rate varies in a dynamic range $0.038 \leq R \leq 0.057$ for $0.001 \leq \beta \leq 100$. The maximum reconnection rate $R=0.057$ is attained at $\beta=0.6$. In the low $\beta$ limit and the high $\beta$ limit, $R \rightarrow 0.050$ and $R \rightarrow 0.038$, respectively. The range of reconnection rate obtained here is consistent with previous classical models of fast reconnection, e.g., the Petschek model ($10^{-2} < R < 10^{-1}$: See Petschek 1964, Vasyliunas 1975). 

Though we can calculate for the range $\beta < 0.001$ and $\beta > 100$ without any difficulty, it is seemingly less meaningful because the curve of $R$ shown in figure \ref{fig:rec-rate} almost settles at each terminal value both for the low $\beta$ limit and the high $\beta$ limit. Since the asymptotic behaviors of both limits $\beta \rightarrow 0$ and $\beta \rightarrow \infty$ are easily predicted, the author thinks our result is enough.

\subsection{Boundary condition along the slow shock}

In paper 2, the inflow region was discussed as a boundary problem by using the Grad-Shafranov (G-S) approach in which we solved a second order partial differential equation (the G-S equation) of elliptic type for the magnetic vector potential $A'_1$ of a perturbed field, where the perturbed field is a deviation from the original uniform magnetic field. In order to solve the G-S equation, we need a boundary condition along the slow shock (the locus of the boundary is approximated as $y=0$ when we put the boundary condition in paper 2 because the opening angle of the X-slow shock is very small). We artificially imposed this boundary condition to be analogous with the result of our numerical simulation in paper 1. Thus, an important question as to how we should impose the boundary condition for the slow shock boundary is still open. The answer is that this boundary condition is spontaneously determined by the shock tube problem discussed in this work. The perturbed magnetic vector potential $A'_1$ can be calculated along the edge of the slow shock (see figure \ref{fig:ss-shape}) with respect to the above solutions.

We can obtain the distribution of $A_1'$ along the $x$-axis ($y=0$) from the solution. However, the boundary value we want is the distribution along the slow shock. Here we assume the shape of the slow shock as shown in figure \ref{fig:ss-shape}. In order to cancel the original uniform magnetic field, we add the correction term $-B_0 y$ to the value on $y=0$. From figure \ref{fig:ss-shape}, the locus of the slow shock is denoted by $y=x \tan \theta$ for $0 \leq x \leq x_c$, $y=[(1-x_s) \tan \phi-x_c \tan \theta](x-x_c)/(1-x_c)+x_c \tan \theta$ for $x_c \leq x \leq 1$. Thus, we obtain the distribution of $A_1'$ as $A_1'=B_1 (x_f-x)+B_2 (x_c-x_f)-B_0 x \tan \theta$ for $0 \leq x \leq x_f$, $A_1'=B_2 (x_c-x)-B_0 x \tan \theta$ for $x_f \leq x \leq x_c$, $-B_{y3} (x-x_c)-B_0\{[(1-x_s)\tan \phi-x_c \tan \theta](x-x_c)/(1-x_c)+x_c \tan \theta\}$ for $x_c \leq x \leq 1$. 

We demonstrate the case $\beta=0.2$ which we numerically simulated in paper 1, and compare the present analytic result with the previous numerical result in figure \ref{fig:bc}. The dotted line and the solid line show the simulation result of paper 1 and the above result of this work, respectively. These two lines are qualitatively analogous, but quantitatively somewhat different. We will discuss the difference in the discussion section \ref{sec:bc}.

\section{Summary \& discussion}
\label{sec:sum-dis}

\subsection{Summary}
\label{sec:sum}

We here briefly summarize the structure of the reconnection system obtained in our numerical simulation in paper 1. The feature of the inner region is very similar to fast reconnection regimes (Priest \& Forbes 1986) of which the Petschek model is one member, i.e., we can find the fast-mode rarefaction dominated inflow and X-shaped slow shock. However, the entire structure shows properties of time-dependent reconnection as follows. The reconnection outflow has a finite span which is increasing in proportion to the time from the onset. A V-shaped forward slow shock forms in the vicinity of the spearhead of the outflow. In region 3, plasmas are compressed by the reconnection jet (piston effect), and squeezed plasmas are gushing out to make a vertical outflow. These are properly included in evolutionary process. 

In this paper, the structure of reconnection outflow in the self-similar evolution model has been studied in a shock tube approximation. We have obtained the reconnection rate and the required boundary condition to solve the inflow region from the structure of the reconnection outflow.

\subsection{Self-Similarity of the reconnection outflow}

We have solved the structure of the reconnection outflow for the self-similar evolution model of fast magnetic reconnection. The structure is spontaneously determined by the reconnection system itself as approximated by a kind of shock tube problem. The reconnection outflow is divided into several regions by discontinuities, e.g., forward shock, reverse shock, and contact discontinuity. We solved this shock tube problem by an iterative (Newton-Raphson) method. The moving speeds of these discontinuities are constant in time in our result. Thus, as expected from the well-known shock tube problem, the reconnection outflow extends self-similarly. 

First, we argue for self-similarity of the evolution of the reconnection system in free space. In the discussion of  paper 2, we need an {\it a priori} statement to authorize a self-similar solution for the inflow region so that the reconnection outflow (in other words, the boundary condition along the slow shock) also evolves self-similarly. The validity of this assumption was, however, unclear in previous papers. We must note that because we have found the self-similarly evolving solution for the reconnection outflow, we can ensure self-similar evolution of the entire reconnection system. 

Comparison of the outflow structure with a shock tube model is also argued in Abe \& Hoshino (2001). They obtained a similar structure to our result, but there is a critical difference. The V-shaped discontinuity of the forward shock is identified as an intermediate shock in their result instead of as a slow shock in our case. The author cannot explain the reason for the difference, but notes that their attention is focused on a much earlier stage than our case. The scale of the plasmoid is the order of the current sheet thickness in their case, but we are interested in a much larger scale structure at the self-similar stage. Actually, the numerical result of Ugai 1999 (see figure 9 of that paper) shows a similar structure to our result. The later evolution of the simulation of Abe \& Hoshino (2001) tends also to coincide with our result.

\subsection{$\beta$-dependence of the reconnection rate}

Second, our attention is focused on the $\beta$-dependence of the reconnection rate $R$. In our result, $R$ is almost constant in the range $0.001 \leq \beta \leq 100$ (see figure \ref{fig:rec-rate}). We could foresee this behavior of the reconnection rate from the following intuition. 

We consider the low $\beta$ limit in the asymptotic region. When the plasma $\beta$ at the asymptotic region far outside the current sheet becomes smaller, the Petschek type X-slow shock must be stronger. This is explained as follows. The gas pressure in the up-stream region is very small relative to the magnetic pressure for a low-$\beta$ plasma, while in the down-stream region, the gas pressure should be very high to keep total pressure equilibrium with the up-stream region. Hence, the total pressure equilibrium imposes a large gas pressure jump at the X-slow shock. This leads to a large jump in the mass density. 

Next we consider the high $\beta$ limit in the asymptotic region. Both sides of the Petschek type X-slow shock are filled with gas-pressure dominated plasmas in this case, and balanced with each other mainly by gas pressure. This results in a very small density jump at the slow shock ($X \rightarrow 1$). Note that this does not mean weak shock limit because the jump of the tangential component of velocity is significant ($v_1 \sim 1$ while $v_{xp}=0$). The $\beta$-dependence of the compression ratio $X$ of the X-slow shock is shown in figure \ref{fig:comp} and is consistent with the above discussion. 

The reconnection rate $R$ is defined as $R \equiv -v_{yp} B_{xp}$. First we discuss the low $\beta$ limit. The inflow speed $-v_{yp}$ will increase as the slow shock strengthens ($\beta$ decreases) for the following reason. The ejection speed of the reconnection jet is almost constant ($\sim V_{A0}$) while the mass density of the jet increases as the slow shock strengthens (the compression ratio $X$ increases as $\beta$ decreases). Thus, the inhalation of mass flux toward the slow shock is also strengthened. This results in an increase of $-v_{yp}$ (see figure \ref{fig:vin-Bin}) as $\beta$ decreases because the mass density at the pre-slow shock region is almost constant (though weakly rarefied by the fast-mode rarefaction wave). As the inflow speed $-v_{yp}$ increases, bending of the magnetic field line becomes remarkable, and $B_{xp}$ decreases (see figure \ref{fig:vin-Bin}). We can also understand the result for the high $\beta$ limit in the same way ({\it vice versa} for the high $\beta$ limit). Consequently, the product of $-v_{yp}$ and $B_{xp}$, and hence the reconnection rate $R$ is almost constant independent of $\beta$. 

Dependence of the reconnection rate on the plasma-$\beta$ is also considered in other works, e.g., Ugai \& Kondoh 2001. Their attention is focused on an implicit contribution of $\beta$ via a current-driven anomalous resistivity model. The aim of our study is intrinsically different from theirs. Needless to say, anomalous resistivity is very important to confirm the validity of a localization of resistivity which is necessary to establish fast reconnection (see Yokoyama \& Shibata 1994). We have concentrated on a macroscopic process which can be treated in the ideal MHD regime and paid scant attention to how we should introduce the anomalous resistivity. We should note that we have no definite information on how to make a relevant macroscopic (MHD) model of anomalous resistivity which should be treated in a microscopic regime.

\subsection{Relation with magnetic Reynolds number}
\label{sec:Rey-indep}

One may feel that our result is strange, because, while the reconnection needs electric resistivity, we treat only ideal MHD and do not consider the dependence on the magnetic Reynolds number. We here discuss a regulation process of the energy conversion speed. As discussed in section \ref{sec:model}, we assume the magnetic diffusion speed is fully adjustable with the spontaneous inhalation speed $-v_{yp}$ which is determined by the above shock tube problem. This assumption is valid for the case of sufficiently large electric resistivity (small magnetic Reynolds number) in which the spontaneous inhalation speed is smaller than the maximum diffusion speed. Such a large resistivity may be realized by anomalous resistivity. We need an assumption for the anomalous resistivity because we do not have a convincing model for the mechanism and estimated value of anomalous resistivity (Coppi \& Friedland 1971, Ji et al. 1998, Shinohara et al. 2001). Note that the diffusion speed is adjustable by a change of the diffusion region thickness even if the resistivity is fixed to be constant, but if there is a lower limit for the thickness (e.g., simulation mesh size or the ion Larmor radius), the diffusion speed is bound by an upper limit. 

In paper 1, we suppose the Reynolds number to be 24.5. One may think that this is an incredibly large resistivity. We should be careful of what denotes the length in the definition of the magnetic Reynolds number. Usually speaking, the magnetic Reynolds number according to Spitzer resistivity for typical solar flares is extremely large: 
\begin{equation}
R_m \sim 10^{13} \frac{(L/10^7 \mbox{[m]})(T/10^6 \mbox{[K]})^{3/2}(B/10^{-3} \mbox{[T]})}{n/10^{15} [\mbox{m}^{-3}]}
\end{equation}
where $L$, $T$, $B$ and $n$ are the dimension of the entire system, the temperature, the magnetic field strength and the plasma number density,  respectively. We must note that ``the effective Reynolds number'' which is based not on the dimension of the entire system but on that of the diffusion region ($L \sim 10^0 \mbox{[m]}$: the ion Larmor radius for solar corona) is important. Hence, the effective Reynolds number $R_e$ for Spitzer resistivity is $R_e \sim 10^6$. If anomalous resistivity which is much larger than the Spitzer resistivity (typically assumed as $\sim 10^{5-6} \times $[Spitzer resistivity] in many numerical MHD simulations) takes place, the effective Reynolds number is estimated as $R_e \sim 10^{0-1}$. The magnetic Reynolds number adopted in paper 1 (which is the effective Reynolds number: $R_e=24.5$) may be relevant for the case of anomalous resistivity. 

A localized resistivity is assumed as a model of anomalous resistivity at the center of the reconnection system. We put the thickness of the initial current sheet $D$ as in paper 1. Here, $D$ is supposed to be of the order of the ion Larmor radius, hence, it is close to the minimum value of the thickness. The dimension of the resistive region is set larger than $D$ (in paper 1, we set $2 D$) in order to deal with change of the diffusion region size. Thus, the effective thickness of the diffusion region can vary between $D$ and $2 D$. The maximum diffusion speed is achieved when the thickness becomes $D$. 

The smallest effective magnetic Reynolds number $R_{e min}$ is defined as 
\begin{equation}
R_{e min} \equiv \frac{V_{A0}}{\eta/D}
\end{equation}
where $\eta$ is the magnetic diffusivity. For the value $R_{emin}=24.5$ in paper 1, the maximum diffusion speed into the diffusion region is determined by the Sweet-Parker model (Sweet 1958, Parker 1963): $v_{dmax} \sim V_{A0}/\sqrt{R_{e min}} \sim 0.2$, while the spontaneous inhalation speed determined by the shock tube approximation is $-v_{yp} \sim 0.05$. The diffusion speed does not limit the inflow speed because $-v_{yp} < v_{dmax}$. This is why the reconnection rate of our model does not depend on the magnetic Reynolds number. 

If the Reynolds number is not so small (small resistivity), say, $R_{e min} \geq 400$, the diffusion speed is smaller than the spontaneous inhalation speed, and may limit the inflow speed. In such cases, the reconnection rate will be suppressed by the small diffusion speed and depends on the magnetic Reynolds number in ways discussed in the previous literature. This problem is very interesting; however, detailed discussion for such cases is beyond the scope of this work. 

In this meaning, the value of $R$ plotted in figure \ref{fig:rec-rate} will be the maximum value for spontaneous reconnection as a function of $\beta$ when the resistivity is sufficiently large ($R_{e min} \leq 400$). The system cannot exceed this value by spontaneous inhalation of the inflow plasma without an external injection. 

A similar discussion about spontaneous reconnection was made by Ugai in his series of works (e.g., Ugai \& Tsuda 1979 and Ugai 1983). He obtained a result supporting Petschek's prediction, i.e., $R \propto (\log R_e)^{-1}$. We here compare with the result of Ugai 1983. His interest is focused on the steady state in a finite simulation box (several times the initial current sheet thickness). This is realized in a very late stage from the onset. At that stage, waves emitted from the central region propagate beyond the entire simulation box (Ugai paid attention to a stage later than roughly 7 times the Alfv\'{e}n transit-time with respect to the entire size of the simulation region). We should note that this is completely different from our case. His resultant reconnection rate logarithmically varies in a range of 0.1-0.3 for $2.5 \leq R_e \leq 200$, and the inflow speed varies in a range of 0.1-0.3 $V_{A0}$ while the maximum diffusion speed (defined by the minimum mesh size) varies between $20 \geq v_{dmax} \geq 0.3$. The author cannot explain why such a very large inflow speed is attained (e.g., inflow speed $-v_{yp} \sim 0.05$ in our case), but we can see that their inflow speed is not so small compared with the maximum diffusion speed for $R_e > 10^2$. Thus, the inflow speed might be restricted by the diffusion speed and the reconnection rate might depend on the magnetic Reynolds number in Ugai's result. The author thinks such a situation is unnatural for astrophysical reconnection as discussed our works (e.g., section \ref{sec:Int} of this paper). Since there is no essential difference between Ugai's numerical procedure and our procedure, if one set a sufficiently wider simulation box than Ugai, he would obtain a similar result to ours.

\subsection{Boundary condition along the slow shock}
\label{sec:bc}

Finally, we discuss the boundary condition for the inflow region which remained as an open question in paper 2. We obtain the distribution of magnetic vector potential $A_1'$ for the perturbed magnetic field along the edge of the reconnection outflow (see figure \ref{fig:bc}). 

The result is roughly consistent with the result of our numerical simulation, but it cannot explain the numerical result quantitatively. The major difference is in the region $x > x_f$. This is mainly due to the two-dimensional effects of the actual outflow. We here assume a quasi-one-dimensional model for the reconnection outflow for simplicity, but the actual outflow clearly has a two-dimensional structure. For example, the magnetic field lines in this region have a round-shaped upward convexity. If we make a precise model of the reconnection outflow and take account of this two-dimensional shape, such a discrepancy will be reduced. Note that, in our experience, the physical quantities in region 1 or 2 which are important to determine the reconnection rate are insensitive to the model of the slow shock shape, and so we do not need to be too concerned about this problem. Obviously, such further discussion cannot proceed without the help of a computer simulation, and is beyond the scope of this work.

\subsection{Relation to stationary models}
\label{sec:stat}

As noted in summary section \ref{sec:sum}, the central region of the entire expanding system tends to be a stationary state which is similar to the Petschek model. The author thinks that the Petschek-like stationary state is unique as the inner solution of spontaneously evolving reconnection systems in free-space with a locally enhanced resistivity (see paper 1). Even when an external circumstance strongly influences the evolution, such the Petschek-like central structure will hold for a long duration of the order of a hundred times Alfv\'{e}n transit time with respect to the proper scale of the external circumstance (e.g., diameter of a magnetic flux tube in case of solar flares). This is because the induced inflow speed is of the order of $10^{-2} V_{A0}$ (see paper 1), and it needs a long time to change the inflow region according to the influence from the external circumstance (see paper 1). 

Other types of stationary states may occur in the following situations: 1) very fast plasma flow ($\sim V_{A0}$: similar to or faster than the FRW propagation) is injected through the boundary, 2) proper scale of the external circumstance is not much larger than the initial current sheet thickness (e.g., in case of laboratory plasma), 3) resistive region is not so localized, etc. However, our self-similar evolution model may be applicable to many cases in astrophysics. \\

The author would like to thank Syuniti Tanuma (Kyoto University), Kazunari Shibata (Kyoto University) and Kiyoshi Maezawa (Japan Aerospace Exploration Agency), who are collaborators on a series of previous papers, for their fruitful comments; Takahiro Kudoh (University of Western Ontario) who gave me a lot of opportunities for useful discussion at the initial stage of this work even in my dream and successively encouraged me. Discussions with anonymous referee were very helpful to improve this paper to be more persuasive. My thanks are due to Mike Kryshak (SOKENDAI) for valuable advice on technical expression in English. I also thank Naoko Kato (SOKENDAI) for her advice on English usage and successive encouragement.

\appendix
\section{Equations for structure of the reconnection outflow}

The structure of the reconnection outflow is determined by the following equations. \\

We assume that the region between the asymptotic region and the pre-shock region is filled with non-resistive plasma. Hence, the magnetic flux is frozen into the induced inflow. We also assume a polytropic variation in the induced inflow because there is no violent process in the fast-mode rarefaction. 
Thus, we impose\\

\noindent
Frozen-in condition [eq1]
\begin{equation}
\frac{B_0}{\rho_0}=\frac{B_{xp}}{\rho_p}
\end{equation}

and \\

\noindent
Polytropic relation [eq2]
\begin{equation}
P_0 \rho_0^{-\gamma}=P_p \rho_p^{-\gamma} \ .
\end{equation}
where $\gamma$ is the specific heat ratio. We assume $\gamma=5/3$ (monoatomic ideal gas).

There are several discontinuities in the reconnection outflow, i.e., X-shaped slow shock, reverse fast shock, contact discontinuity, and forward V-shaped slow shock (see figure \ref{fig:sch}). We set jump conditions for both sides of each discontinuity:\\

\noindent
X-shaped slow shock Rankine-Hugoniot (R-H) jump conditions\\

\noindent
Pressure-jump [eq3]
\begin{eqnarray}
\frac{P_1}{P_p}=&1+\frac{\gamma}{c_{sp}^2}(\cos \theta v_{yp})^2 (X-1) \times
\{
\frac{1}{X}-(\cos \theta B_{xp}+\sin \theta B_{yp})^2/2 \nonumber \\
&\times [-2 V_{Apx}^2 X+(\cos \theta v_{yp})^2 (X+1)]/[((\cos \theta v_{yp})^2-X V_{Apx}^2)^2 \mu \rho_p]
\}
\end{eqnarray}

where
$$c_{sp}=\sqrt{\gamma P_p/\rho_p} \ ,$$ 
$$V_{Apx}=\sqrt{(-\sin \theta B_{xp}+\cos \theta B_{yp})^2/(\mu \rho_p)} \ ,$$
$\mu$ is the magnetic permeability of vacuum and $X$ is the compression ratio. \\

\noindent
Density-jump [eq4]
\begin{equation}
\frac{\rho_1}{\rho_p}=X
\end{equation}

\noindent
Velocity (parallel) jump [eq5]
\begin{equation}
\frac{\cos \theta v_1-v_0}{\sin \theta v_{yp}-v_0}=\frac{(\cos \theta v_{yp})^2-V_{Apx}^2}{(\cos \theta v_{yp}^2-X V_{Apx}^2)}
\end{equation}

where $v_0=v_{yp} B_{xp}/(B_{xp} \sin \theta-B_{yp} \cos \theta)$ is the shift speed of the de Hoffmann-Teller coordinate.

\noindent
Velocity (perpendicular) jump [eq6]
\begin{equation}
\frac{-\sin \theta v_1}{\cos \theta y_{yp}}=\frac{1}{X}
\end{equation}

\noindent
Magnetic field (parallel) jump [eq7]
\begin{equation}
\frac{\sin \theta B_1}{\cos \theta B_{xp}+\sin \theta B_{yp}}=\frac{[(\cos \theta v_{yp})^2-V_{Apx}^2]X}{(\cos \theta v_{yp})^2-X V_{Apx}^2}
\end{equation}

\noindent
Magnetic field (perpendicular) jump [eq8]
\begin{equation}
\frac{\cos \theta B_1}{-\sin \theta B_{xp}+\cos \theta B_{yp}}=1
\end{equation}

The compression ratio $X$ is defined by the following equation (3rd order algebraic equation for $X$),
\begin{eqnarray}
&\{
(-\sin \theta B_{xp}+\cos \theta B_{yp})^2 [(\cos \theta B_{xp}+\sin \theta B_{yp})^2 (\gamma - 1) \rho_p (\cos \theta v_{yp})^2 \nonumber \\
&+(-\sin \theta B_{yp}+\cos \theta B_{yp})^2 (2 \gamma P_p-\rho_p (\cos \theta v_{yp})^2+\gamma \rho_p (\cos \theta v_{yp})^2)]
\} X^3 \nonumber \\
&+
\{
-\rho_p (\cos \theta v_{yp})^2 
[
(-\sin \theta B_{xp}+\cos \theta B_{yp})^4 (\gamma-1) \nonumber \\
&+
(\cos \theta B_{xp}+\sin \theta B_{yp})^2 (\gamma-2) \mu \rho_p (\cos \theta v_{yp})^2 \nonumber \\
&+
(-\sin \theta B_{xp}+\cos \theta B_{yp})^2
(
(\cos \theta B_{xp}+\sin \theta B_{yp})^2 (\gamma+1) \nonumber \\
&+
2 \mu (2 \gamma P_p-\rho_p (\cos \theta) v_{yp})^2+\gamma \rho_p (\cos \theta v_{yp})^2
)
]
\} X^2 \nonumber \\
&+
\{
\mu \rho_p (\cos \theta v_{yp}^4
[
(\cos \theta B_{xp}+\sin \theta B_{yp})^2 \gamma+
2(-\sin \theta B_{xp}+\cos \theta B_{yp})^2 (\gamma+1)+
\mu \nonumber \\
&(
2 \gamma P_p-\rho_p (\cos \theta v_{yp})^2+\gamma \rho_p (\cos \theta v_{yp})^2
)
]
\} X \nonumber \\
&+
\{
-(\gamma+1) \mu^2 \rho_p^3 (\cos \theta v_{yp})^6
\}=0 \ . \nonumber
\end{eqnarray} 
This equation has three roots. We must choose a real positive root larger than unity.

\noindent
Reverse-fast shock R-H conditions\\

\noindent
Pressure-jump [eq9]
\begin{equation}
\frac{P_2}{P_1}=\zeta_f
\end{equation}

where 
$$\zeta_f=\gamma M_{1f}^2 (1-1/\xi_f)+(1-\xi_f^2)/\beta_1+1$$ 
with 
$$\xi_f=(-l + \sqrt{l^2 + 2/\beta_1 (2 - \gamma) (\gamma + 1) \gamma M_{1f}^2})/(2/\beta_1 (2 - \gamma)) \ ,$$
$$l = \gamma (1/\beta_1 + 1) + (\gamma - 1) \gamma M_{1f}^2/2 \ ,$$
$$\beta_1 = P_1/(B_1^2/(2 \mu)) \ ,$$
$$M_{1f} = (v_1 - x_f V_A)/c_f \ ,$$
$$c_f = \sqrt{\gamma P_1/\rho_1}$$
and 
$$V_A = B_0/\sqrt{\mu \rho_0}\ .$$

\noindent
Density-jump [eq10]
\begin{equation}
\frac{\rho_2}{\rho_1}=\xi_f
\end{equation}

\noindent
Velocity jump [eq11]
\begin{equation}
\frac{v_1-x_f}{v_2-x_f}=\xi_f
\end{equation}

\noindent
Magnetic field jump [eq12]
\begin{equation}
\frac{B_2}{B_1}=\xi_f
\end{equation}

We must impose local magnetic flux conservation on both sides of region p and region 1. \\

\noindent
Magnetic flux conservation at X-point [eq13]
\begin{equation}
v_{yp} B_{xp}+v_1 B_1=0
\end{equation}

\noindent
Force balance at contact discontinuity [eq14]
\begin{equation}
P_2+\frac{B_2^2}{2 \mu}=P_3+\frac{B_{x3}^2+B_{y3}^2}{2 \mu}
\end{equation}

\noindent
V-shaped forward slow shock R-H conditions\\

\noindent
Pressure-jump [eq15]
\begin{equation}
P_3=P_0 \left\{
1+\frac{\gamma}{c_{s0}^2} (x_s \sin \phi)^2 (X_h-1)
\times \left[
\frac{1}{X_h}-\frac{(B_0 \cos \phi)^2}{2} \frac{-2 V_{A0x}^2 X_h+(x_s \sin \phi)^2 (X_h+1)}{[(x_s \sin \phi)^2-X_h V_{A0x}^2] \mu \rho_0}
\right]
\right\}
\end{equation}
where
$c_{s0}=\sqrt{\gamma P_0/\rho_0}$, $V_{A0x}=\sqrt{(-\sin \theta B_0)^2/(\mu \rho_0)}$ and $X_h$ is the compression ratio. \\

\noindent
Density-jump [eq16]
\begin{equation}
\frac{\rho_3}{\rho_0}=X_h
\end{equation}

\noindent
Velocity (parallel) jump [eq17]
\begin{equation}
\frac{\cos \phi (v_{x3}-x_s V_{A0})+\sin \phi v_{y3}}{-V_{A0} x_s \cos \phi}
=\frac{v_{m0x}^2-V_{A0x}^2}{v_{m0x}^2-X_h V_{A0x}^2}
\end{equation}
where $v_{m0x}=x_s \sin \phi$ and $V_{A0}=B_0/\sqrt{\mu \rho_0}$.

\noindent
Velocity (perpendicular) jump [eq18]
\begin{equation}
\frac{-\sin\phi (v_{x3}-x_s V_{A0})+\cos \phi v_{y3}}{V_{A0} x_s \sin \phi}=\frac{1}{X_h}
\end{equation}

\noindent
Magnetic field (parallel) jump [eq19]
\begin{equation}
\frac{\cos \phi B_{x3}+\sin \phi B_{y3}}{B_0 \cos \phi}=\frac{(v_{m0x}^2-V_{A0x}^2) X_h}{v_{m0x}^2-X_h V_{A0x}^2}
\end{equation}

\noindent
Magnetic field (perpendicular) jump [eq20]
\begin{equation}
\frac{-\sin \phi B_{x3}+\cos \phi B_{y3}}{-B_0 \sin \phi}=1
\end{equation}

The compression ratio $X_h$ is a solution of the following third order algebraic equation:
\begin{eqnarray}
&\{
(B_0 \sin \phi)^2 
[
(B_0 \sin \phi)^2 (\gamma-1) \rho_0 (x_s \sin \phi)^2+
(B_0 \sin \phi)^2 
(
2 \gamma P_0-\rho_0 (x_s \sin \phi)^2+\gamma \rho_0 (x_s \sin \phi)^2
)
]
\} X_h^3 \nonumber \\
&+
\{
-\rho_0 (x_s \sin \phi)^2
[
(B_0 \sin \phi)^4 (\gamma+1)+(B_0 \cos \phi)^2 (\gamma-2) \mu \rho_0 (x_s \sin \phi)^2 \nonumber \\
&+(B_0 \sin \phi)^2 
(
(B_0 \cos \phi)^2 (\gamma+1)+(2 \mu (2 \gamma P_0-\rho_0 (x_s \sin \phi)^2+\gamma \rho_0 (x_s \sin \phi)^2))
)
]
\} X_h^2 \nonumber \\
&+
\{
\mu \rho_0^2 (x_s \sin \phi)^4 
(
(B_0 \cos \phi)^2 \gamma + 2(B_0 \sin \phi)^2 (\gamma+1)+
\mu (2 \gamma P_0 - \rho_0 (x_s \sin \phi)^2+\gamma \rho_0 (x_s \sin \phi)^2)
)
\} X_h \nonumber \\
&+
\{
-(\gamma+1) \mu^2 \rho_0^3 (x_s \sin \phi)^6
\}=0\ . \nonumber
\end{eqnarray}

We must choose a real positive root larger than unity. 

The tip of the reconnection outflow touches the FRWF (see figure \ref{fig:ss-shape}), hence $A_1'=0$ at this point. This condition reduces to the following equation. 

\noindent
Boundary Condition at the tip of the outflow [eq21]
\begin{equation}
-B_{y3} (1-x_c) + B_{x3} [(1-x_s) \tan \phi - x_c \tan \theta]-B_0 (1-x_s) \tan \phi =0
\end{equation}

From magnetic flux conservation, the injected magnetic flux must be redistributed in the reconnection jet. This leads to the following equation. 

\noindent
Magnetic flux conservation [eq22]
\begin{equation}
v_{yp} B_{xp}+[x_f B_1+ (x_c-x_f) B_2]=0
\end{equation}

We assume the following trivial relations:\\

\noindent
Definition from the contact discontinuity
$$
x_c=v_2=v_{x3} 
$$

We can solve [eq1] for $\rho_p$, [eq2] for $P_p$, [eq8] for $B_1$, [eq9] for $P_2$, [eq10] for $\rho_2$, [eq12] for $B_2$, [eq13] for $v_{yp}$, [eq15] for $P_3$, [eq16] for $\rho_3$, [eq17] and [eq18] for $v_{x3}$ and $v_{y3}$, [eq19] and [eq20] for $B_{x3}$ and $B_{y3}$ by hand, then substitute them into other nine equations ([eq3], [eq4], [eq5], [eq6], [eq7], [eq11], [eq14], [eq21] and [eq22]) for the following unknowns: $v_1$, $B_{xp}$, $B_{yp}$, $\theta$, $P_1$, $\rho_1$, $x_f$, $\phi$ and $x_s$. The only parameter included in this problem is the plasma-$\beta$ value at the asymptotic region. By using a Newton-Raphson routine, with an initial guess for these unknowns, we obtain converged solutions. The procedure to obtain the series of converged solutions is discussed in section \ref{sec:b-eqs} in detail.

\appendix

\figcaption[]{Schematic figure of the reconnection outflow. Several discontinuities form in the outflow. We consider a quasi-one dimensional shock tube problem along the $x-$ and $y-$axes. 
\label{fig:sch}
}

\figcaption[]{$\beta$-dependence of the reconnection rate. The reconnection rate is almost constant ($\sim 0.05$) as a function of plasma-$\beta$ at the asymptotic region far outside the current sheet. The obtained value of the reconnection rate is consistent with previous theoretical and numerical works, but we must note that this value is spontaneously determined by the reconnection system itself in a frame work of ideal (non-resistive) MHD. 
\label{fig:rec-rate}
}

\figcaption[]{Model for the slow shock geometry. We adopt a simplified geometry of the ``crab hand'' shaped slow shock in order to calculate the distribution of $A_1'$ (perturbed magnetic vector potential). 
\label{fig:ss-shape}
}

\figcaption[]{Boundary condition along the slow shock. The perturbed magnetic vector potential $A_1'$ for $\beta=0.2$ is obtained along the slow shock based on the structure of the reconnection outflow (solid line). This is important to determine the structure of the inflow region (see Nitta et al. 2002). The result of our previous simulation is also shown by dotted line. These two are analogous, but we can find a difference especially around the contact discontinuity ($x=0.87$). This is a two dimensional effect of the actual outflow, and it shows the limit of our one dimensional approximation and simplified geometry of the slow shock. We must note that the value of the reconnection rate (crossing point with $x=0$) is in good agreement with the simulation result. 
\label{fig:bc}
}

\figcaption[]{Compression ratio of X-slow shock. Compression ratio $X$ of the Petschek model-like X-slow shock near $x=0$ is plotted as a function of plasma-$\beta$ at the asymptotic region. $X$ increases as $\beta$ decreases. 
\label{fig:comp}
}

\figcaption[]{Inflow speed and inflow magnetic field strength. In order to understand the behavior of the reconnection rate shown in figure \ref{fig:rec-rate}, we plot the two factors (the inflow speed $-v_{yp}$ and the inflow magnetic field $B_{xp}$) of the reconnection rate. As $\beta$ decreases, the X-shock strength increases (see discussion of this paper). This leads to the behavior of $-v_{yp}$ and $B_{xp}$ shown in this figure, and {\it vice versa} for the high $\beta$ limit. Thus, the reconnection rate which is the product of these two quantities is almost constant independent of $\beta$. 
\label{fig:vin-Bin}
}


\begin{thebibliography}{99}
\baselineskip=1.0pc

\bibitem[Abe 2001]{AH}
Abe, S.A., Hoshino, M., 2001, Earth Planets Space, 53, 663

\bibitem[Coppi \& Friedland 1971]{Cop71}
Coppi,B., \& Friedland,A.B., 1971, ApJ, 169, 379

\bibitem[Hayashi et al. 1999]{H99}
Hayashi,M.R., Shibata,K., Matsumoto,R., 1999, proc of ``Star formation 1999'', P.288

\bibitem[Ji et al. 1998]{Ji98}
Ji,H., Yamada,M., Hsu,S., \& Kulsrud,R., 1998, Phys. Rev. Lett., 80, 3256

\bibitem[Koyama et al. 1986]{YSO-K}
Koyama,K., Maeda,Y., Ozaki,M., Ueno,S., Kamata,Y., Tawara,K., Skinner, S. \& Yamauchi,S., 1994, PASJ, 46, L125

\bibitem[Koyama et al. 1986]{GRXE-K}
Koyama,K., Makishima,K., Tanaka,Y., \& Tsunemi,H., 1986, PASJ, 38, 121

\bibitem[Nitta et al. 2001]{Nit01}
Nitta, S., Tanuma, S., Shibata, S., Maezawa, K., 2001, ApJ, 550, 1119

\bibitem[Nitta et al. 2002]{Nit02}
Nitta, S., Tanuma, S., Maezawa, K., 2001, ApJ, 580, 538

\bibitem[Parker 1963]{Par}
Parker, E. N. 1963, ApJ Suppl. Ser., 8,177

\bibitem[Petschek 1964]{Pet}
Petschek,H.E. 1964, NASA Spec. Publ., 50, AAS-NASA Symposium on Physics of Solar Flares, 425

\bibitem[Priest \& Forbes 1986]{P-F}
Priest,E.R. \& Forbes,T.G. 1986, J.Geophys.Res, 91, 5579

\bibitem[Sato \& Hayashi 1979]{S-H}
Sato,T. \& Hayashi,T. 1979, Phys.Fluids, 22, 1189

\bibitem[Shinohara et al. 1998]{SNFTMTY1998}
Shinohara, I., Nagai, T., Fujimoto, M., Terasawa, T., Mukai, T., Tsuruda, K. \& Yamamoto, T., 1998, JGR, 103, 20365 

\bibitem[Sweet 1958]{Swe}
Sweet, P. A. 1958, The neutral point theory of solar flares, in Electro-magnetic Phenomena in Cosmical Physics, ed. B.Lehnert (Cambridge University Press, London), 123

\bibitem[Tanuma et al. 1999]{T99}
Tanuma,S., Yokoyama,T., Kudoh,T., Matsumoto,R., Shibata,K.\& Makishima,K., 1999, PASJ, 51, 161

\bibitem[Ugai 1983]{U83}
Ugai,M. 1983, Phys.Fluids, 26, 1569

\bibitem[Ugai \& Kondoh 2001]{U-K}
Ugai,M. \& Kondoh,K., 2001, Phys. Plasmas, 8, 1545

\bibitem[Ugai \& Tsuda 1977]{U-T1}
Ugai,M. \& Tsuda,T. 1977, J. Plasma Phys., 17, 337

\bibitem[Ugai \& Tsuda 1979]{U-T2}
Ugai,M. \& Tsuda,T. 1979, J. Plasma Phys., 22, 1

\bibitem[Ugai 1999]{Uga2}
Ugai,M. 1999, Phys. Plasmas, 6, 1522

\bibitem[Vasyliunas 1975]{Vas}
Vasyliunas,V.M. 1975, Rev.Geophys., 13, 303

\bibitem[Yokoyama \& Shibata 1994]{Y-S}
Yokoyama, T. \& Shibata, K. 1994, ApJ, 436, L197

\end{thebibliography}
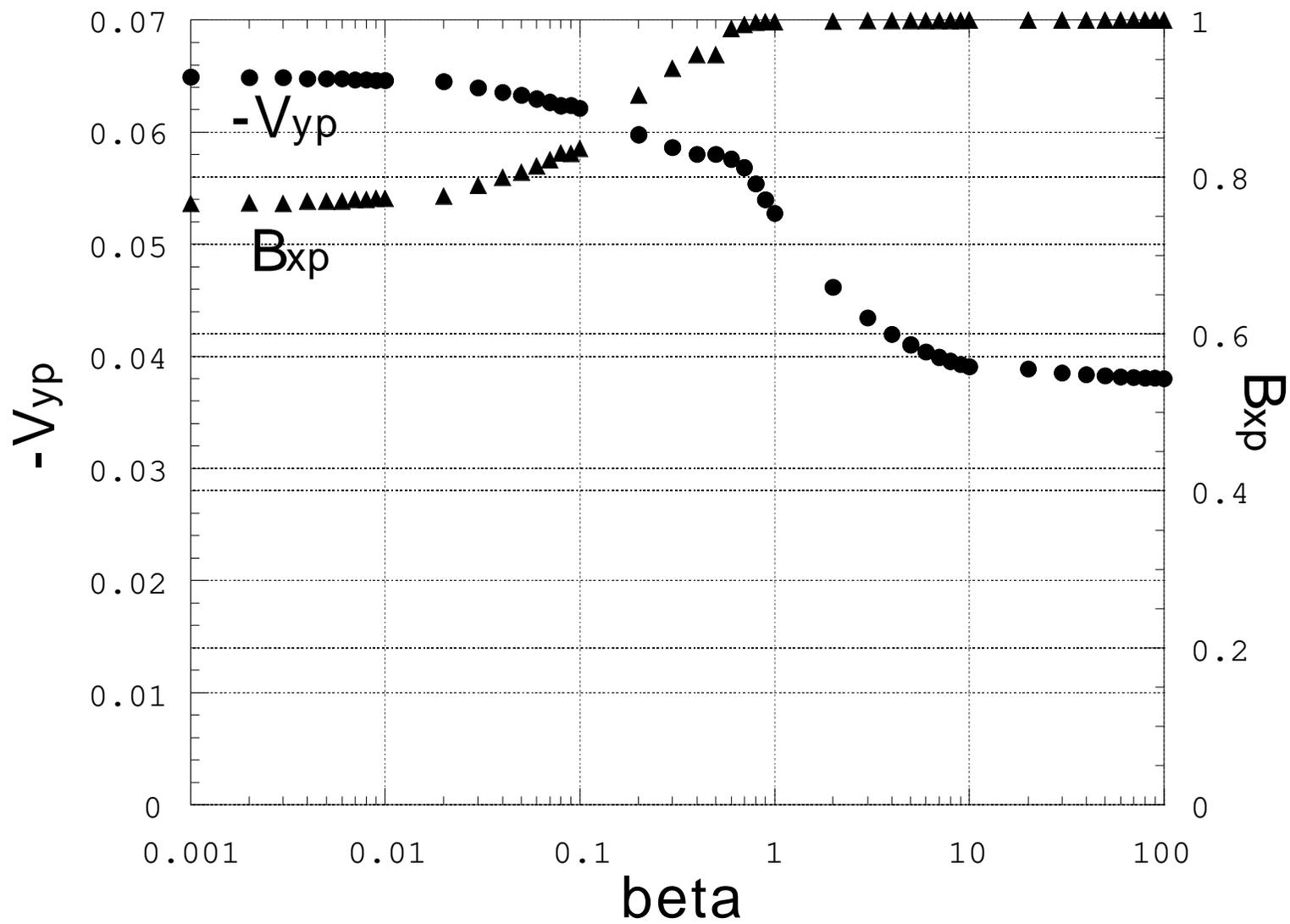
\end{document}